\documentclass[aps,prl,twocolumn,floatfix,superscriptaddress,showpacs,amsmath,amssymb]{revtex4}
\usepackage{graphicx}
\usepackage{epsfig}  
\usepackage{xcolor}
\allowdisplaybreaks

\newcommand{\be}{\begin{equation}}
\newcommand{\ee}{\end{equation}}
\newcommand{\bea}{\begin{eqnarray}}
\newcommand{\eea}{\end{eqnarray}}
\newcommand{\ba}{\begin{eqnarray*}}
\newcommand{\ea}{\end{eqnarray*}}

\newcommand{\bR}{\mathbf{R}}

\newcommand{\bk}{\mathbf{k}}

\newcommand{\bRp}{\mathbf{R'}}

\newcommand{\m}[1]{\mathcal{#1}} 

\newcommand{\spin}[1]{\sigma^{#1}}

\newcommand{\omq}{\omega_{q}} 	
\newcommand{\omr}{\omega_{r}} 	
\newcommand{\omo}{\omega_{0}} 	

\newcommand{\Eq}[1]{Eq.~(\ref{#1})}

\newcommand{\Fig}[1]{Fig.~\ref{#1}}

\begin{document}
 
\title{On the phase transition of light in cavity QED lattices}
\author{M. Schir\'o}
\affiliation{Princeton
 Center for Theoretical Science and Department of Physics, Joseph Henry
  Laboratories, Princeton University, Princeton, NJ
  08544, USA}
\author{M. Bordyuh}
\affiliation{Department of Electrical Engineering, Princeton University, Princeton, NJ
  08544, USA}
 \author{B. \"{O}ztop}
 \affiliation{Institute for Quantum Electronics, ETH-Z\"urich, CH-8093 Z\"urich,
Switzerland}
\affiliation{Department of Electrical Engineering, Princeton University, Princeton, NJ
  08544, USA}
  \author{H. E. T\"ureci}
\affiliation{Department of Electrical Engineering, Princeton University, Princeton, NJ
  08544, USA}
\affiliation{Institute for Quantum Electronics, ETH-Z\"urich, CH-8093 Z\"urich,
Switzerland}  

\date{\today} 
 
\pacs{42.50 -o, 42.50 Pq, 73.43.Nq, 05.30.Rt}
\begin{abstract} 
Systems of strongly interacting atoms and photons, that can be realized wiring up individual cavity QED systems into lattices, are perceived as a new platform for quantum simulation. While sharing important properties with other systems of interacting quantum particles, here we argue that the nature of light-matter interaction gives rise to unique features with no analogs in condensed matter or atomic physics setups. By discussing the physics of a lattice model of delocalized photons coupled locally with two-level systems through the elementary light-matter interaction described by the Rabi model, we argue that the inclusion of counter rotating terms, so far neglected, is crucial to stabilize finite-density quantum phases of correlated photons out of the vacuum, with no need for an artificially engineered chemical potential. We show that the competition between photon delocalization and Rabi non-linearity drives the system across a novel $Z_2$ parity symmetry-breaking quantum criticality between two gapped phases which shares similarities with the Dicke transition of quantum optics and the Ising critical point of quantum magnetism. We discuss the phase diagram as well as the low-energy excitation spectrum and present analytic estimates for critical quantities.
\end{abstract}
\maketitle

\textit{Introduction -}  Interaction between light and matter is one of the most basic processes in nature and represents a cornerstone in our understanding of a broad range of physical phenomena. In the study of strongly correlated systems and collective phenomena, light has traditionally assumed the role of a spectroscopic probe. The increasing level of control over light-matter interactions with atomic and solid-state systems~\cite{haroche_exploring_2006, yamamoto_semiconductor_2000, schoelkopf_wiring_2008} has brought forth a new class of quantum many-body systems realized on photon lattices~\cite{hartmann_strongly_2006, greentree_quantum_2006, angelakis_photon-blockade-induced_2007, rossini_mott-insulating_2007, aichhorn_quantum_2008, na_strongly_2008, carusotto_fermionized_2009, schmidt_strong_2009, koch_superfluidmott-insulator_2009, pippan_excitation_2009, tomadin_signatures_2010,
umucallarartificial_2011} where light and matter play equally important roles in emergent phenomena.

 The basic building block of such systems is the elementary Cavity QED (CQED) system formed by a two-level system (TLS) interacting with a single mode of an electromagnetic resonator.
When CQED systems are coupled to form a lattice, the interplay between photon blockade \cite{tian_quantum_1992, imamoglu_strongly_1997, rebic_large_1999} and inter-cavity photon tunnelling leads to phenomenology akin to those of Hubbard models of massive bosons as realized e.g. by ultracold atoms in optical lattices \cite{bloch_many-body_2008}. The possibility of  quantum phase transitions of light between Mott-like insulating and superfluid phases has stimulated a great deal of discussion recently \cite{hartmann_strongly_2006, greentree_quantum_2006, angelakis_photon-blockade-induced_2007, rossini_mott-insulating_2007, aichhorn_quantum_2008, na_strongly_2008, schmidt_strong_2009, koch_superfluidmott-insulator_2009, pippan_excitation_2009, tomadin_signatures_2010}. The excitement about these systems stems from their potential as dissipative quantum simulators that provide full access to individual sites through continuous weak measurements \cite{houck_on-chip_2012}. 
 
While sharing important features with conventional condensed matter or atomic physics setups, systems of strongly correlated photons have their own unique properties that ultimately derive from the nature of the fundamental light-matter interaction. As photons can disappear by interacting with the matter field, their number is not conserved but rather fixed by the condition of thermal equilibrium. To describe this situation for a photon gas in equilibrium with either photonic or dipolar bath -- such as in a blackbody -- one says that photons have zero chemical potential~\cite{landau_statistical_1980}.  
From the point of view of bosonic Hubbard models 
this has rather dramatic consequences, as one would then require an external non-equilibrium drive in order to engineer non-trivial quantum many body states other than the vacuum \cite{klaers_bose-einstein_2010}. For a Lattice CQED system however, as we will show in this Letter, this is remarkably not so. The light-matter interaction strength can play the role of an effective chemical potential to stabilize finite density quantum phases of correlated photons out of vacuum. We find that the so-called counterclockwise terms in the elementary CQED Hamiltonian, the Rabi model, typically neglected for quantum optical systems are relevant perturbations that explicitly break the conservation of total number of excitations. This changes completely the nature of quantum criticality of lattice CQED systems from those based on Jaynes-Cummings model~\cite{hartmann_strongly_2006, greentree_quantum_2006, angelakis_photon-blockade-induced_2007, rossini_mott-insulating_2007, aichhorn_quantum_2008, na_strongly_2008, carusotto_fermionized_2009, schmidt_strong_2009, koch_superfluidmott-insulator_2009, pippan_excitation_2009, tomadin_signatures_2010,
umucallarartificial_2011}, leading to a $Z_2$ parity-breaking quantum phase transition where the two level systems polarize to generate a ferroelectrically ordered state and the photon coherence acquires a non-vanishing expectation value due to hopping. This novel quantum criticality, described by a delocalized super-radiant quantum critical point, shares some similarity with the Dicke phase transition of quantum optics and turns out to be in the universality class of the Ising model. 

\textit{Single Resonator - } The elementary light-matter interaction between a photonic mode of a resonator and a TLS is described by the Rabi model~\cite{haroche_exploring_2006}
\be\label{eqn:Hrabi}
\m{H}_{R}= \omr \,a^{\dagger}\,a + \omq\sigma^+\,\sigma^-+
g\,x\,\sigma_x 
\ee
where $\omr$ is the frequency of the resonator, $\omq$ the qubit transition frequency, $g$ the light-matter coupling strength and $x=a+a^{\dagger}$. 
In addition, depending on the specific context, an extra term should be added to Eq.~(\ref{eqn:Hrabi}) where the field appears quadratically, $H_{A^2}=D\,(a+a^{\dagger})^2$. We will discuss its implications at the end of the Letter and first address the physics of the Rabi model~(\ref{eqn:Hrabi}) and its lattice extension.
When the coupling $g$ is sufficiently smaller than the frequencies $\omr$, $\omq$ one can safely neglect processes where simultaneously atomic and photonic excitations are created, described by the counter-rotating terms $a^{\dagger}\,\sigma^+ + \sigma^-\,a$. In this so called rotating-wave approximation the Hamiltonian~(\ref{eqn:Hrabi}) reduces to the Jaynes-Cumming (JC) model, $H_{JC}=
\omr\,a^{\dagger}\,a+\omq\sigma^+\,\sigma^-+g\left(a^{\dagger}\,\sigma^-+\sigma^+\,a\right)$ used widely in discussions of CQED physics. While appropriate in many relevant cases, recent implementations of circuit QED \cite{niemczyk_circuit_2010, forn-diaz_observation_2010} achieved coupling strengths $g$ where the counter-rotating terms begin to show significant deviations from the expectations of the JC model~\cite{hussin_ladder_2005, haroche_exploring_2006, larson_dynamics_2007, hausinger_dissipative_2008, werlang_rabi_2008, bourassa_ultrastrong_2009, casanova_deep_2010, beaudoin_dissipation_2011}. This is the so called ``ultra-strong coupling'' regime of parameters where $g$ becomes a considerable fraction of $\omega_r$.
 
\begin{figure}[t]
\begin{center}
\epsfig{figure=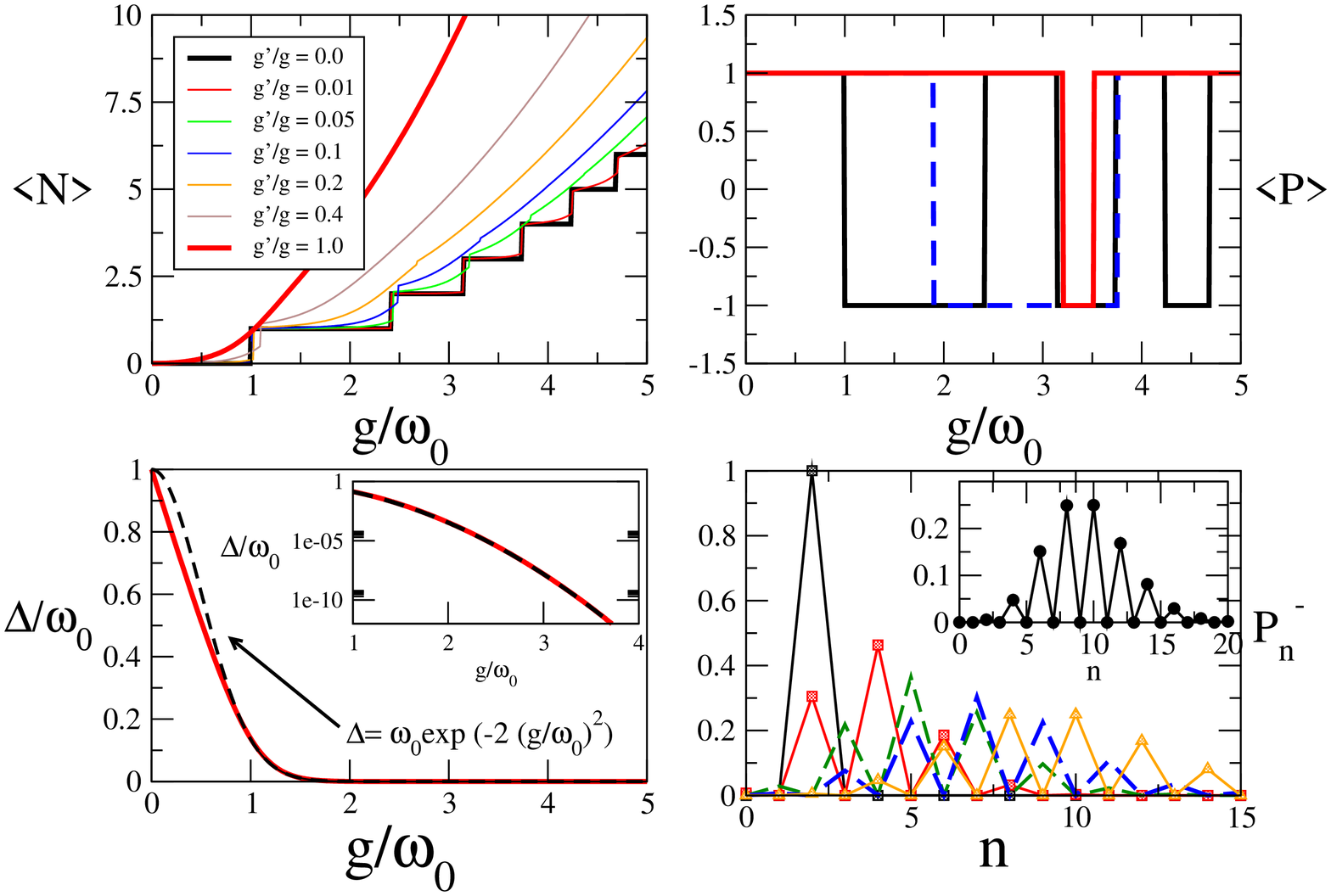,scale=0.32}
\caption{Evolution of the single site generalized Rabi Model ground state properties upon increasing the strength of counter-rotating terms $g'/g$. Top panels: number of excitations and parity of the ground state as a function of $g/\omo$ and for different $g'/g$. Bottom panels: (left) decay of level splitting $\Delta$ in the ultra-strong coupling regime and (right) probability of having $n$ polaritons in the Rabi ground state (inset) and its evolution with $g'/g=0.0,0.25,0.5,0.75,1.0$.}
\label{fig:singlesite}
\end{center}
\end{figure}

Although the physics of the Rabi model has been widely studied and well-understood \cite{graham_two-state_1984, kus_exact_1986, reik_exact_1986}
it is attracting renewed attention recently~\cite{braak_integrability_2011, solano_dialogue_2011}.
Here we will examine the Rabi-Hubbard model as realized e.g. by a lattice of circuit QED cavities where each node is described by $\m{H}_R$.
As we discuss below, this system forms a viable platform for studying non-trivial strongly correlated phases of light. Experimental efforts to fabricate on-chip photonic lattices of circuit QED systems are currently underway \cite{Andrew2012}.
Before we introduce the lattice, we consider the generalized Rabi model
\bea\label{eqn:g_Rabi}
\m{H}_{gR} [a,a^{\dagger}]= \omr\,a^{\dagger}\,a+\omq\sigma^+\,\sigma^-+
g\,\left(a^{\dagger}\,\sigma^-+\sigma^+\,a\right)+\nonumber\\
+g'\,\left(a^{\dagger}\,\spin{+}\,+\,\spin{-}\,a\right)
\eea
We would like to stress that we introduce this model to explore the role of counter-rotating terms in a controlled fashion. This model interpolates between the JC Hamiltonian for $g'=0$ and the standard Rabi Hamiltonian for $g'=g$. In the following we will restrict ourselves to the resonant case $\omr=\omq=\omo$. For $g'=0$ i.e. in the JC limit, the above model conserves the total number of excitations, $\m{N}=a^{\dagger}\,a+\spin{+}\spin{-}$. The resulting continuous $U(1)$ symmetry allows an exact analytic solution of $\m{H}_{gR}$	 in terms of dressed states of photons and TLS excitations, the polaritons. The ground-state shows an interesting evolution upon increasing the coupling $g/\omo$, with an infinite series of level crossings for $g_c(n)= \omo\,\left(\sqrt{n+1}+\sqrt{n}\right)$ where the number of excitations increases from $n$ to $n+1$, resulting in a characteristic staircase structure (see \Fig{fig:singlesite}). 

For an arbitrarily small $g'$, counter-rotating terms break the continuous $U(1)$ symmetry down to a discrete $Z_2$ group associated with parity $\m{P}=e^{i\pi\m{N}}$. Under this unitary operator the photon field $a$ and the TLS operator $\sigma_x$ transform respectively as $\m{P}^{\dagger}a\m{P}=-a$ and $\m{P}^{\dagger}\sigma_x\m{P}=-\sigma_x$, from which the invariance immediately follows, namely $[\m{P},\m{H}_{gR}]=0$. A direct consequence of the parity symmetry is that, while $\langle\,a\rangle=\langle\,\sigma_x\rangle=0$ in the ground state of $\m{H}_{gR}$ much as in the JC limit, the photon field in the  Rabi ground state is squeezed, i.e. $\langle\,a^{2n}\rangle\neq0$.
While the discrete $Z_2$ symmetry prevent a full closed-form solution, the model~(\ref{eqn:g_Rabi}) in the Rabi limit $g'=g$ has recently been shown to be nevertheless integrable
~\cite{braak_integrability_2011}. Important features of this exact solution that will be relevant for our discussion below are that (i) no level crossing between states of different parities can occur as a function of $g/\omo$ (note that $g'=g$), which in turn implies that the ground state of the Rabi model remains an even parity state for any $g/\omo$, (ii) the ground state and the first excited state are quasi-degenerate in the ultra-strong coupling regime $g\gtrsim\omo$.

To get further insight into the structure of the Rabi ground state 
we numerically diagonalize the Hamiltonian~(\ref{eqn:g_Rabi}). In \Fig{fig:singlesite} (top panels) we plot the number of excitations and the parity of the ground state as a function of $g/\omo$ for different values of $g'/g$. Upon increasing the strength of counter-rotating terms the JC plateaux are gradually smeared out. Though the parity remains well-defined, the evolution with $g'$ reveals multiple crossings between eigenstates switching the parity of the ground state, ultimately resulting in an even parity ground state when $g'=g$. 
We also plot (bottom panel, \Fig{fig:singlesite}) the scaled level splitting $\Delta/\omo$ between the ground state and the first excited state, which vanishes as $\Delta\sim e^{-2\left(g/\omo\right)^2}$ for large $g/\omega_0$ in agreement with degenerate perturbation theory. In contrast, the gap to the next energy level stays of order one (not shown) at large $g/\omo$. Polaritonic dressed states are not anymore exact eigenstates for $g'\neq0$. Turning on $g'$ results in a broadening of the eigenstates of the generalized Rabi model when projected on the polaritonic eigenstates of the JC model, its peak shifting as a function of $g'$ resulting in a ground state containing a finite number of excitations (right panel of ~\Fig{fig:singlesite}) .

\textit{Lattice model of interacting atoms and photons -- } We now come to the main subject of this Letter, which is the physics of the Rabi-Hubbard model, a model of itinerant photons hopping between neighboring resonators and interacting on-site with a TLS according to the local Hamiltonian (\ref{eqn:g_Rabi}). The full many-body Hamiltonian for this system reads
\be\label{eqn:HgR_lattice}
\m{H} = -J\,\sum_{\langle \bR\bRp\rangle}\,\,a^{\dagger}_\bR\,a_\bRp+\sum_{\bR}\,\m{H}_{gR}[a_\bR,\sigma_\bR^+] .
\ee
We stress here that, with an eye on possible future experiments on a circuit QED architecture, we do not include any chemical potential to tune the density of excitations in the ground state. The goal here is to exploit the spontaneous polarization of the Rabi vacuum that emerges when either the light-matter interaction strength $g$ or the hopping strength $J$ is increased.  

We start considering first the $g'=0$ JC limit and study the phase diagram in the $g-J$ plane~\cite{JC}. 
In the absence of hopping, the ground state is an exact dressed state of polaritons. A gapped and incompressible Mott Insulating (MI) phase of polaritons survives at finite hopping until a critical value of $J$ is reached. The phase boundary $J_c(g)$ features characteristic Mott lobes (see \Fig{fig:fig3}), a legacy of the level crossings of the single site JC model discussed above. For hopping strengths larger than $J_c$ the system is in a superfluid (SF) compressible phase with gapless excitations associated to phase fluctuations of the $U(1)$ order parameter. 
It is now well-established that the JC lattice model is in the same universality class as the Bose-Hubbard model \cite{aichhorn_quantum_2008, koch_superfluidmott-insulator_2009, schmidt_strong_2009, pippan_excitation_2009, tomadin_many-body_2010, hohenadler_dynamical_2011}.
Crucially for our purpose here, the experimental realization of this MI-SF quantum phase transition requires an external driving or a suitably engineered chemical potential in order to counter-balance photon losses into the vacuum.

\begin{figure}[t]
\begin{center}
\epsfig{figure=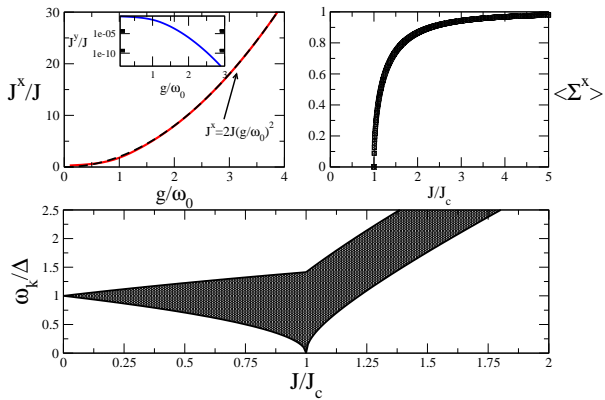,scale=0.3}
\caption{Top panels: (Left) Scaled ferromagnetic coupling $J^{x,y}/J$ of the effective spin model as a function of $g/\omo$. While $J^x\sim (g/\omo)^2$ at ultra-strong coupling, the coupling along $y$ is vanishingly small. (Right) Mean-field order parameter for the effective spin model. Bottom panel: Band of spin-wave excitations above the ground-state. 
}
\label{fig:fig2}
\end{center}
\end{figure}

\begin{figure}[t]
\begin{center}
\epsfig{figure=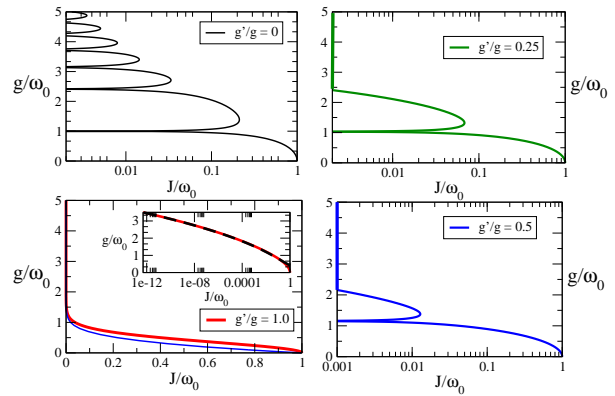,scale=0.3}
\caption{Mean field phase diagram for the generalize Rabi lattice model for different values of $g'/g$. 
For the Rabi case, $g'=g$, we compare with the mean field phase diagram for the effective spin model (left bottom panel). In the inset, we show the decay of the critical coupling $J_c$ at ultra-strong coupling $J_c=\omo^3\,e^{-2\left(g/\omo\right)^2}/4g^2$~\cite{caveat}} 
\label{fig:fig3}
\end{center}
\end{figure}

We now argue that the inclusion of counter-rotating terms in \Eq{eqn:HgR_lattice} has a dramatic effect on the above physics. 
The roots of this can be traced back to the single resonator limit. As discussed above, the counter-rotating terms leave the system with a discrete $Z_2$ symmetry associated to parity. Photon hopping in (\ref{eqn:HgR_lattice}) can trigger a spontaneous breaking of this parity symmetry above some critical coupling $J_c(g)$, toward a phase where both $\langle\,a_\bR\rangle\neq0$ and $\langle\,\sigma^x_\bR\,\rangle\neq0$. As the broken symmetry is discrete, this quantum phase transition is fundamentally different from the JC one. 
Indeed it can be seen as a delocalized super-radiant quantum critical point reminiscent of the multi-mode Dicke transition of quantum optics.
In order to see that a non-zero $J$ favors ordering, we start from the full Hamiltonian~(\ref{eqn:HgR_lattice}) and notice that photons 
can be integrated out exactly in an imaginary-time action formalism to obtain an effective model for the TLSs only. The result of this calculation~\cite{rabi_si} reveal that photon mediates an effective Ising-like coupling between TLS which is retarded and long-range,  $J^{eff}_{\bR-\bRp}(\tau)=-g^2/2\,\langle\,T_{\tau}x_{\bR}(\tau)\,x_{\bRp}(0)\rangle$. The scaling with $g$ implies that at sufficiently large $g/\omo$ and for finite $J$, a ferromagnetically ordered Ising phase emerges with $\langle\sigma^x_\bR\,\rangle\neq0$.
Further insight into this emerging $Z_2$ degree of freedom are obtained from the single site limit.
As we discussed, at large $g/\omo$ the ground state and the first excited state are almost degenerate, with an exponentially small splitting and a gap to the next level which stays of order one.
These two states $\vert\pm\rangle$ have opposite parity and can be thought as eigenstates of an effective pseudospin $1/2$ degree of freedom, $\Sigma^z_{\bR}$.
 In addition we notice that the photon operator $a_{\bR}$ does not couple states with same parity. Its expression in the restricted $\vert\pm\rangle$ subspace reads $a_{\bR}\rightarrow\,\beta\,\Sigma_{\bR}^++\gamma\,\Sigma_{\bR}^-$, where the dependence of the coefficients $\beta$, $\gamma$ on the coupling $g$ can be obtained numerically. In the limit $g/\omo\gg1$ one can analytically show that a linear scaling holds $\beta=\gamma\sim g/\omo$. 
  Armed with these results we can rewrite the Rabi-Hubbard Hamiltonian as 
\bea\label{eqn:Heff}
\m{H}_{eff} = -\sum_{\langle\bR\bR'\rangle} J^{x}\,\Sigma^x_{\bR}\,\Sigma^x_{\bR'}
+J^{y}\,\Sigma^y_{\bR}\,\Sigma^y_{\bR'}
+\frac{\Delta}{2}\sum_{\bR}\,\Sigma^z_{\bR}
\eea
where the couplings $J^{x,y}=J(\gamma\pm\beta)^2/2$ depend on $g$ as shown in \Fig{fig:fig2}. 
This effective model describes a pseudospin anisotropic $XY$ model in a longitudinal magnetic field $\Delta/2$, which is known to display a quantum phase transition toward a $Z_2$ broken symmetry phase which is in the Ising universality class for any finite anisotropy, $J^{x}\neq J^{y}$~\cite{sachdev_quantum_2011}. 
The effective psuedospin description highlights once more the differences between the Rabi and the JC case. Indeed here both the disordered and the ordered phases are gapped except right at the critical point where the gap is expected to vanish as a power-law.

\textit{Mean Field Phase Diagram and Fluctuations - }  We now use Gutzwiller mean field theory to confirm the general picture we have drawn for the transition. By decoupling the hopping term in~(\ref{eqn:HgR_lattice}) we reduce the original lattice problem to an effective single site problem, $H_{eff}[\psi]=H_{loc}-\psi\,\left(a^{\dagger}+a\right)$ in a self-consistent field $\psi=Z\,J\langle\,a\rangle_{\psi}$.
By expanding the energy to second order in $\psi$ we can get the mean field phase boundary $J_c(g)$, above which a parity symmetry broken phase emerges with both $\langle\,a_{\bR}\rangle\neq0$ and $\langle\,\sigma_{\bR}^x\rangle\neq0$.
In \Fig{fig:fig3} we plot the mean field phase boundary in the $J,g$ plane for different values of $g'/g$ from the JC to the Rabi limit. The "Mott lobes" for $g'=0$ are gradually suppressed as the ratio $g'/g$ is increased. For intermediate values a residual lobe structure remains, which reflects the level crossings already discussed in the single-site problem. However we stress that no Mott insulator exists for any finite $g'$.
Further insight on the transition can be gained from the effective spin Hamiltonian~(\ref{eqn:Heff}). A linearized fluctuation analysis gives a critical coupling $J_c=\Delta/\left(\beta+\gamma\right)^2\sim \omo^3\,e^{-2\left(g/\omo\right)^2}/4g^2$  which agrees with the numerical results in the large $g/\omo$ regime (see figure ~\ref{fig:fig3}). In addition the effective Hamiltonian also gives access to the spectrum of low-lying excitations $\omega_{\bk}$, plotted in \Fig{fig:fig2}, that as expected is gapped on both sides and vanishes in a power-law fashion at the transition, $E_g\sim\vert J-J_c\vert^{1/2}$. 
The low-energy spectrum is gapless at the critical point, with a linear dispersion $\omega_{\bk}=c\,\vert\bk\vert$.

\textit{Discussion - } The physical picture we have drawn from our analysis of the lattice Rabi model reveals a striking feature of hybrid systems made of atoms and photons. Due to the nature of the fundamental light-matter interaction, which allows non-trivial vacuum fluctuations, no external driving forces or artificially engineered chemical potentials are in principle required to stabilize  finite density quantum phases of correlated atoms and photons. 
Rather it is the coupling between matter and light that will trigger this non-trivial vacuum polarization. An important question for a possible experimental realization e.g. on a circuit QED platform concerns the stability of the above picture against photon leakage that is an inherent feature of any quantum optical system.  Physical intuition would suggest that at least for a small coupling to a low-temperature photonic bath the ordered phase would be protected by the discrete nature of the $Z_2$ symmetry. However an in-depth study of the phase diagram and a full understanding of quantum criticality in the open system limit is an important fundamental problem that we leave to future investigation. 

We now briefly discuss the role of a generic $A^2$ term $H_{A^2}=D\,(a+a^{\dagger})^2$ in the picture that emerges from the above discussion. Recently it has been argued that such a term, generally assumed to be small, can become relevant in certain cavity QED realizations of the \emph{single-mode} Dicke Model, where an ensemble of many TLSs is coupled to a single mode of a cavity. Indeed when the coupling $D$ scales sufficiently fast with light matter interaction $g$, $D>g^2/\omega_q$, the super-radiant critical point disappears. While this condition is realized in cavity QED setups with real atoms coupled via electric dipole and results in so called no-go theorems, the situation with some circuit QED implementations, where TLSs couple capacitively to the resonator, is currently subject of scientific debate~\cite{VonDelft_nogo,NatafCiuti_NatComm}. 
We note that in contrast to the single-mode Dicke model, in our system the $Z_2$ parity symmetry breaking emerges from a non-trivial interplay between delocalization of photons through hopping
and local light-matter interactions. As a result the critical boundary $J_c(g)$ can be accessed by increasing the hopping strength $J$, at fixed (and even moderate) light-matter coupling. While the inclusion of $H_{A^2}$ in our lattice Hamiltonian may change quantitatively the shape of the phase boundary~\cite{rabi_si}, especially in the ultrastrong coupling regime, it is rather an issue of the specific implementation that will determine the ideal architecture to realize the Rabi phase transition in an experimental system. Finally we note that circuit QED implementations can be engineered where the $A^2$ term is irrelevant. This is the case, for example, of flux qubits inductively-coupled to resonators~\cite{NatafCiuti_PRL}, a setup that in principle~\cite{DevoretGirvinSchoelkopf} is ideally suited to access the ultra-strong coupling regime and that has been recently explored experimentally in single circuit QED units~\cite{niemczyk_circuit_2010}.

\textit{Conclusions - } In this work we have explored the physics of itinerant photons hopping between neighboring resonators of a lattice of CQED systems. We have studied its equilibrium phase diagram as a function of the atom-photon coupling $g$ and shown that this system displays a novel parity symmetry breaking quantum phase transition, belonging to the $Z_2$ Ising universality class, between two gapped phases. Simultaneously, the photonic degrees of freedom acquire a non-vanishing expectation value, displaying a delocalized superradiant phase above a critical hopping.

\textit{Note added - } The following reference~\cite{Zheng_Takada_PRA} that is relevant to the discussion in this manuscript was brought to our attention after the publication of this work.

\textit{Acknowledgements -} We thank David Huse for valuable and stimulating discussions. This work was supported by the National Science Foundation  through the Princeton Center for Complex Materials under grant no.\ DMR-0819860 and by the Swiss NSF through grant no.\  PP00P2-123519/1.

\bibliographystyle{apsrev}


\begin{thebibliography}{45}
\bibitem[{pga()}]{caveat}
\bibinfo{journal}{We expect the Gutzwiller mean field results to be accurate in the regime $J\ll\omega_0$. 
The behaviour at large hoppings $J\sim\omega_0$ requires a different approach as the tight-binding model (3) itself becomes questionable in this regime, 
where a significant mixing of single-resonator modes occur.}
\end{thebibliography}

\appendix

\section{Supplementary Material for ''On the phase transition of light in cavity QED lattices''}


\subsection{Integrating out the Photons: Effective Spin-Spin Interactions }

An interesting perspective on the physics of a model of localized spins and itinerant photons can be obtained by integrating out the photonic degrees of freedom. This can be done exactly as the photons always enter quadratically. Let us first rewrite the quadratic part of the action associated to the Rabi model (we consider this case for simplicity, extension to the generalized Rabi is straightforward). After diagonalizing the hopping in Fourier space we get
\bea
S_0 = \int\,d\tau \sum_{\bk} a_\bk(\tau)\left(\partial_{\tau}-\omega_{\bk}\right)\bar{a}_{\bk}(\tau)=\\
\int\,d\tau \sum_{\bk} a_\bk(\tau)\,\m{G}^{-1}_{\bk}(\tau-\tau')\bar{a}_{\bk}(\tau')
\eea
where the photonic dispersion reads $ \omega_{\bk}=\Omega-2J\sum_{\alpha}\,cos\,k_{\alpha}$ and we have introduced the bare photon Green's function
\be
\m{G}_{\bk}(\tau)=-\langle\,T_{\tau}\,a_{\bk}(\tau)\,a^{\dagger}_{\bk}(0)\rangle 
\ee
The coupling between the photon and the spin is linear in the field $x_i=a_i+\bar{a}_i$
\be
S_g = g\,\int_0^{\beta}\,d\tau\, \sum_i\,x_i(\tau)\,\sigma^x_i(\tau) 
\ee
which after Fourier transforming reads
\bea
S_g = g\,\int_0^{\beta}\,d\tau\, \sum_{\bk}\,x_{\bk}(\tau)\,\sigma^x_{-\bk}(\tau) =\nonumber\\
=g\,\int_0^{\beta}\,d\tau\, \sum_{\bk}\,a_{\bk}(\tau)\,\sigma^x_{-\bk}(\tau) + a^{\dagger}_{\bk}(\tau)\,\sigma^x_{\bk}(\tau) \,.
\eea
%
The integration over the photon field can be done exactly by a simple shift of variable and the result reads
\be
Z=\int\,D\,a\,D\,\bar{a}\,D\sigma\,e^{-\left(S_0+S_g+S_{loc}\right)}= \int\,D\sigma\,e^{-S_{eff}}
\ee
with a spin-only effective action of the form
\be
S_{eff}= S_{loc}+g^2\,\int\,d\tau\,d\tau'\,\sum_{\bk} \,\sigma^x_{-\bk}(\tau)\,\m{G}_{\bk}(\tau-\tau')\,
\sigma^x_{\bk}(\tau') 
\ee
where  $\m{S}_{loc} = \omega_q\, \sum_{i} \sigma^z_{i}$ is proportional to the total inversion of the TLSs, which acts as a transverse field. We can rewrite this effective action as following 
\bea
 S_{eff}=\m{S}_{loc}+
\int\,d\tau\,d\tau'\,\sum_{\bk} \,\sigma^x_{-\bk}(\tau)\,J_{\bk}\left(\tau-\tau'\right)
\sigma^x_{\bk}(\tau')
\eea
where  the effective spin-spin interaction is given in terms of the Green's function of the photon field
\be
J_{\bk}(\tau)=\frac{g^2}{2}\left(\m{G}_{\bk}(\tau)+\m{G}_{\bk}(-\tau)\right)
\ee
We notice this coincides with the Matsubara Green's function of the photon displacement
\be
J_{\bk}(\tau) = -\frac{g^2}{2}\,\langle\,T_{\tau}\,x_{\bk}(\tau)\,x_{-\bk}(0)\rangle
\ee
a result which is also expected by looking at the order by order perturbation theory. 

\section{The Rabi Lattice model with the $A^2$ term}

In this section we discuss the modification of the phase boundary of the Rabi Lattice model when the $A^2$ term of the form
\be 
H_{A^2}=D\,\sum_{\bR}\,(a_{\bR}+a^{\dagger}_{\bR})^2\,
\ee
is added. For the sake of simplicity we consider the case $g'=g$, hence the full Hamiltonian reads
\bea\label{eqn:new_rabi}
\m{H} = -J\,\sum_{\langle \bR\bRp\rangle}\,\,a^{\dagger}_\bR\,a_\bRp+\sum_{\bR}\,\m{H}_{R}[a_\bR,\sigma_\bR^+]+H_{A^2}
\eea
where $\m{H}_R$ is the hamiltonian of the single site Rabi model. We start noticing that, at a general level, this extra term does not break explicitly the $Z_2$ symmetry of the problem, hence it doesn't \emph{a priori} rule out the existence of the phase transition. At the same time we expect this term to modify the phase boundary in such a way to favour the disordered (symmetric) phase.

We notice at this point that, contrarily to the case of single mode Dicke Model, our lattice Hamiltonian cannot be solved exactly in general.  In order to proceed it is useful to perform a canonical transformation on the photon fields to eliminate $H_{A^2}$. This can be done by expressing the photon operators $a_{\bR},a^{\dagger}_{\bR}$on each lattice site in terms of a new set of degrees of freedom $b_{\bR},b^{\dagger}_{\bR}$
\begin{eqnarray}
a_{\bR} = cosh\theta\,b_{\bR}+sinh\theta\,b^{\dagger}_{\bR}\\
a^{\dagger}_{\bR} = sinh\theta\,b_{\bR}+cosh\theta\,b^{\dagger}_{\bR}
\end{eqnarray}
By choosing 
$$ tanh\,2\theta = -2D/\left(\omega_r+2D\right) 
$$
we can eliminate the $A^2$ term and obtain a local Hamiltonian which is again of the Rabi form, yet with renormalized photon parameters
\be
\tilde{H}_{rabi} = \sum_{\bR}
\tilde{\omega}_r\,b^{\dagger}_{\bR}\,b_{\bR}+\frac{\omega_q}{2}\,\sigma^z_{\bR}\,+
\tilde{g}\,\sigma^x_{\bR}\,(b^{\dagger}_{\bR}+b_{\bR}) 
\ee
where 
$$
\tilde{\omega}_r=\sqrt{\omega_r^2+4\,D\,\omega_r}\,\qquad\,
\tilde{g}=g\sqrt{\omega_r/\tilde{\omega}_r}
$$
 We further notice that the hopping term between different resonators gets also renormalized and in addition a pair-hopping term arises so that the hopping Hamiltonian reads in terms of the new fields
\bea
\tilde{H}_{hopping}= -\tilde{J}\,\sum_{\langle \bR\bRp\rangle}\,\left(b^{\dagger}_\bR\,b_\bRp+hc\right)+\\
-\tilde{J}_{pair}\,\sum_{\langle \bR\bRp\rangle}\,\left(b_\bR\,b_\bRp+b^{\dagger}_{\bR}\,b^{\dagger}_{\bRp}\right)
\eea
where the renormalized hopping read respectively
\be\label{eqn:hoppings}
\tilde{J}=J\,\frac{\omega_r+2D}{\tilde{\omega}_r}\,\qquad\,
\tilde{J}_{pair}=-J\,\frac{2D}{\tilde{\omega}_r}\,\qquad\,
\ee
We start noticing that, quite generically for $D>0$, the pair hopping term satisfies the condition
\be
\frac{\vert \tilde{J}_{pair}\vert}{\tilde{\omega}_r}=\left(\frac{2D/\omega_r}{1+4\,D/\omega_r}\right)\,\frac{J}{\omega_r}<\frac{J}{\omega_r}
\ee
from which we conclude that if we assumed in the original problem $J\lesssim\omega_r$ (an assumption which is behind the Hamiltonian~\ref{eqn:new_rabi}) then, within the same accuracy, we can drop the pair hopping terms in the transformed hamiltonian. 
\begin{figure}[h]
\begin{center}
\epsfig{figure=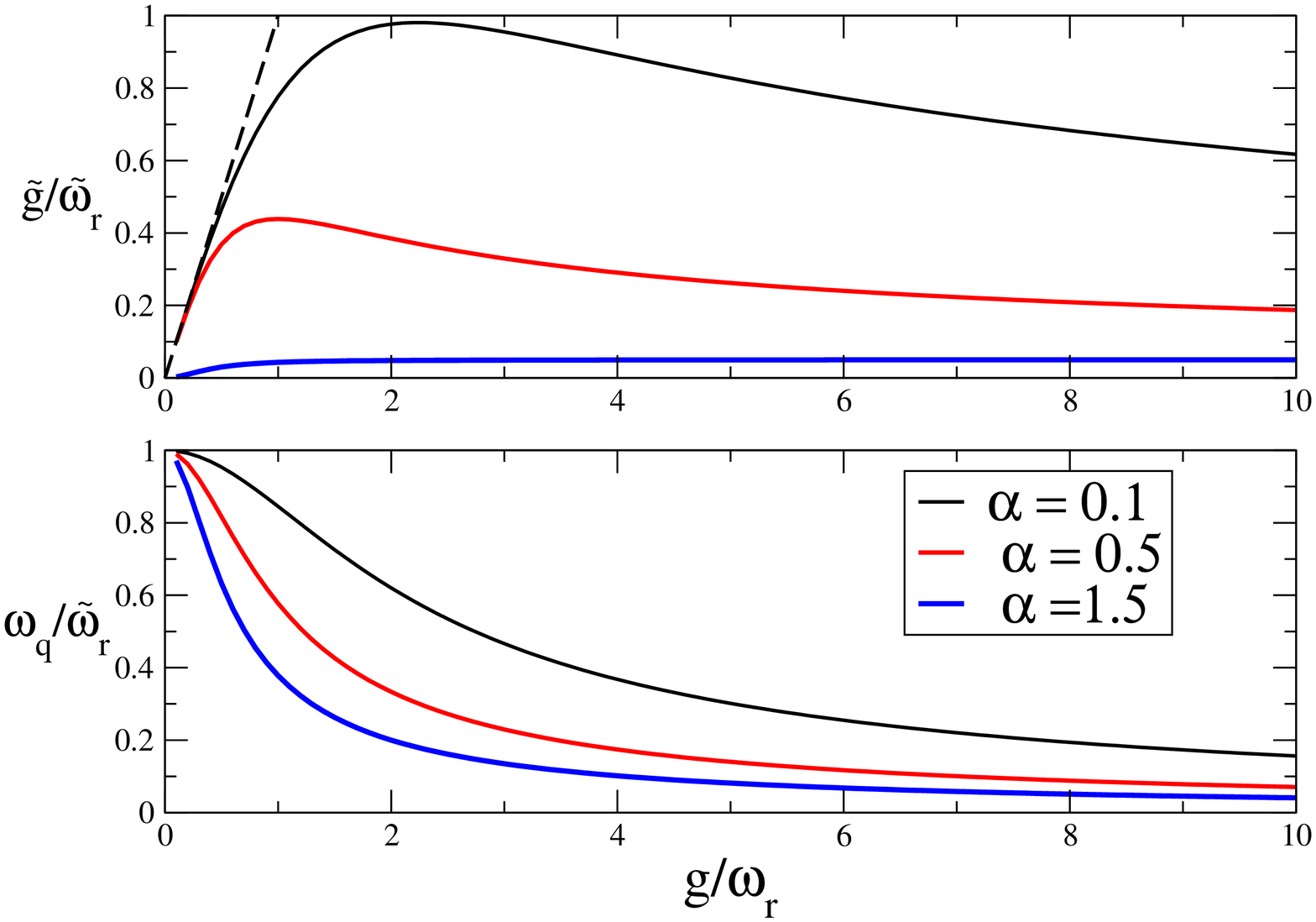,scale=0.3}
\epsfig{figure=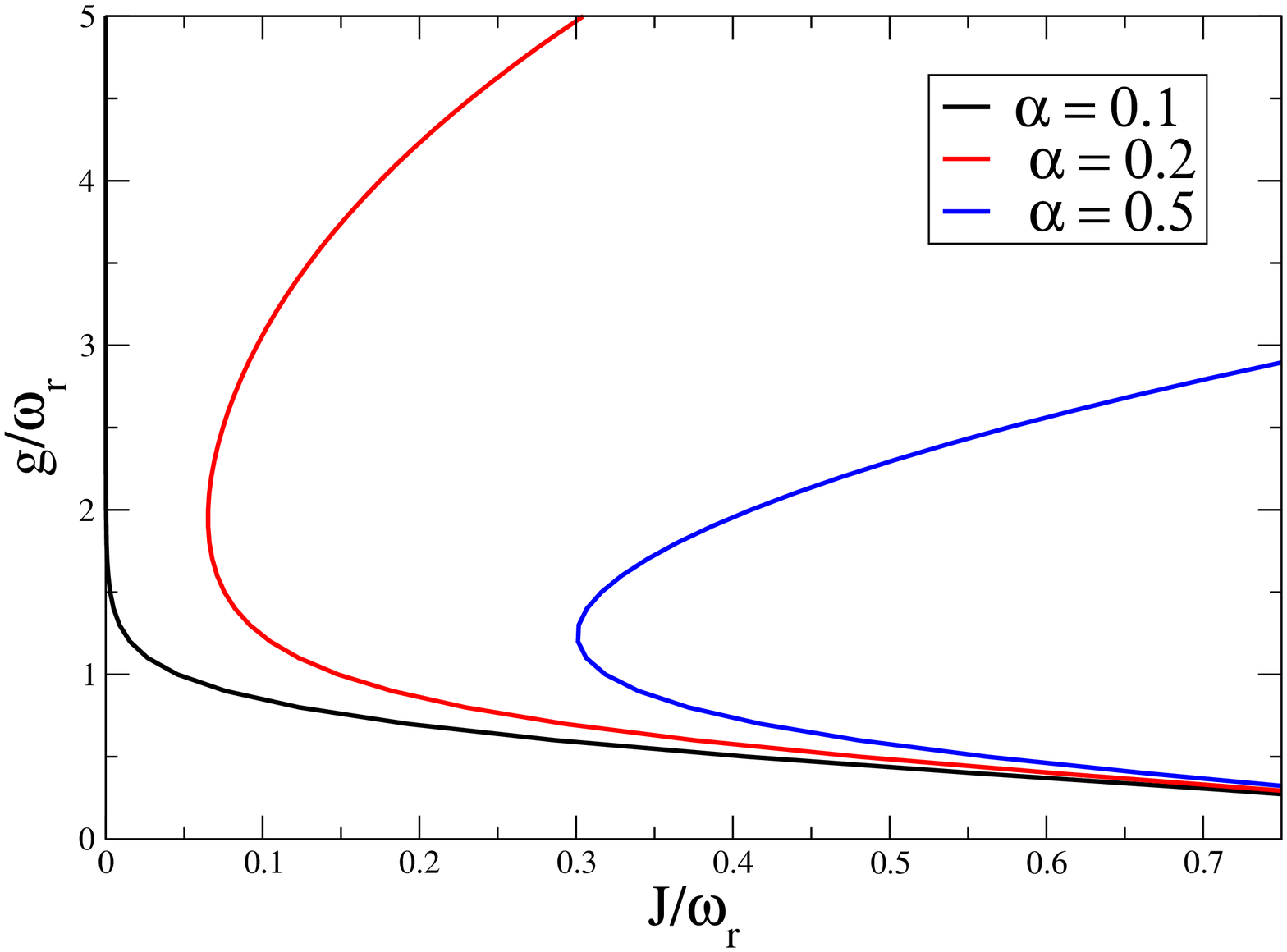,scale=0.3}
\caption{Left Panel: Renormalized light-matter coupling $\tilde{g}/\tilde{\omega}_r$ and TLS frequency $\omega_q/\tilde{\omega}_r$ as a function of the bare coupling $g/\omega_r$ for different values of the parameter $\alpha$. We notice that for small $g/\omega_r$ the effects of renormalization are negligible for any $\alpha$ and become more relevant for larger values of $g/\omega_r$. Right Panel: Mean Field Phase diagram for different values of the parameter $\alpha$.}\label{fig:renormalized}
\label{fig:fig1}
\end{center}
\end{figure}
From the above discussion we conclude that the effect of the $A^2$ term can be fully encoded in a new lattice Rabi model with renormalized parameters. 
Whenever the coupling $D$ can be made independent of the light-matter coupling $g$, a situation that can arise in certain circuit QED architectures, the effect of the $A^2$ term is just a trivial uniform shift of the phase boundary in the $g$ vs $J$ plane. The opposite situation is more tricky. We will assume in the following a coupling $D$ that depends on the light-matter interaction $g$ as
$D=\alpha\frac{g^2}{\omega_q}$~and discuss the physics of the model for different values of $\alpha$.  In order to gain insight, it is useful to look at how the coupling $D$ renormalizes the photon parameters. In Figure \ref{fig:renormalized} we plot for different values of $\alpha$ the renormalized effective light-matter coupling $\tilde{g}/\tilde{\omega}_r$ and the renormalized effective TLS frequency $\omega_q/\tilde{\omega}_r$ as a function of $g/\omega_r$.  We see that in general the role of a finite $\alpha$ is (i) to scale down the effective light-matter interaction particularly in the ultrastrong coupling regime $g/\omega_r\gg1$, and (ii) to renormalize down the TLS frequency $\omega_q/\tilde{\omega}_r\ll1$, toward a regime of large detuning. We notice that  the two effects compete with each other, as the former reduces the tendency to order while the latter enhances it. This competition, in the case of \emph{single-mode} Dicke model, has a rather drastic consequence for sufficiently large $\alpha$ -- the disappearance of the superradiant critical point. 
However, while the phase transition in the Dicke model is driven only by the local physics, namely g vs. $\omega_q$, in the Rabi model increasing the hopping strength $J$ is another route to access the super-radiant phase. For this we find that the hopping strength is only weakly affected by the renormalization due to $D$. It is easy to see using Eq.~(\ref{eqn:hoppings}) and the definition of $D$ that
$\tilde{J}/\tilde{\omega}_r\sim J/\omega_r$, even at ultra-strong coupling.
In conclusion, we expect the critical point to be accessible even in presence of the $A^2$ term. While a thorough discussion of this issue will be presented elsewhere, here we give support to this claim by solving the Rabi Hubbard model with the $A^2$ term within the Gutzwiller mean field approximation As we see from the right panel of Figure~\ref{fig:renormalized} the shape of the phase boundary is only quantitatively affected by finite $\alpha$,  particularly in the ultra-strong coupling regime, but the overall picture is not changed, namely a hopping driven quantum phase transition between a disordered normal phase and a delocalized super-radiant phase where the TLSs order ferromagnetically.

\end{document}